\documentclass[10pt]{iopart}
\usepackage{graphicx}
\usepackage{xcolor}
\usepackage{float}
\usepackage{amssymb}

\begin{document}

\title[]
{
Hydrodynamics of a dense flock of sheep: edge motion and long-range correlations
}

%%% author information
\author{Marine de Marcken$^1$ and Rapha\"{e}l Sarfati$^2$ }
\address{$^1$ Center for One Health Research, Department of Environmental and Occupational Health Sciences, University of Washington, Seattle, Washington 98195, USA}
\address{$^2$ Department of Chemical and Biological Engineering, University of Colorado Boulder, Boulder, Colorado 80309, USA}
\ead{raphael.sarfati@colorado.edu}

%%% abstract
\begin{abstract}

Sheep are gregarious animals, and they often aggregate into dense, cohesive flocks, especially under stress.
In this paper, we use image processing tools to analyze a publicly available aerial video showing a dense sheep flock moving under the stimulus of a shepherding dog.
Inspired by the fluidity of the motion, we implement a hydrodynamics approach, extracting %density and 
velocity fields, and measuring their propagation and correlations in space and time.
We find that while the flock overall is stationary, significant dynamics happens at the edges, notably in the form of fluctuations propagating like waves, and large-scale correlations spanning the entire flock.
These observations highlight the importance of incorporating interfacial dynamics, for instance in the form of line tension, when using a hydrodynamics framework to model the dynamics of dense, non-polarized swarms.

\end{abstract}

%
% Uncomment for keywords
%\vspace{2pc}
\noindent{\it Keywords}: swarming, collective motion, sheep, self-organization, animal behavior, shepherding
%
% Uncomment for Submitted to journal title message
%\submitto{\BB}
%
% Uncomment if a separate title page is required
%\maketitle
% 
% For two-column output uncomment the next line and choose [10pt] rather than [12pt] in the \documentclass declaration
\ioptwocol

\section{Introduction}

Sheep are notoriously gregarious animals.
Their instinct to remain in a close-packed configuration, both at rest or in motion, is common knowledge, and has been repeatedly described in the literature, fictional and  scientific.
Rabelais writes famously, of Panurge's sheep: ``The flock was such that once one jumped, so jumped its companions. It was not possible to stop them, as you know, with sheep, it's natural to always follow the first one, wherever it may go" \cite{Rabelais1552}.
Since 3000~B.C., shepherds have observed that sheep stay in close proximity, and follow the flock faithfully, if not at times recklessly (\textit{ibid.}).
Sheep have also been observed to collectively drift towards places of higher elevations, and to ``close-in'' on themselves under stress \cite{Hamilton1971,King2012,Ginelli2015}.
These characteristics are understood to result from the fragility of an individual sheep against the natural predators, which are many.
By staying in a dense configuration, sheep attempt to lower their odds of being caught.
By assuming a higher position, they have a better point of view to detect incoming predators \cite{Hamilton1971}.

Sheep flocks constitute a particularly interesting experimental system to study the collective behavior of social organisms (called \textit{flocking} or \textit{swarming}) for a number of reasons, including: 
the extent of empirical knowledge about sheep flocking behavior; 
the large availability of domesticated sheep flocks; 
the density of the aggregation, which limits sensory information to immediate vicinity; 
and the fact that sheep motion is purely two-dimensional, making tracking and analysis easier than flocks of birds or schools of fish.
Swarming, from bacteria to wildebeests, has always captivated ethologists, but only recently started to pique the interest of engineers and physicists as well. 
From a fundamental point of view, the emergent large-scale dynamics resulting from a multitude of individual interactions is reminiscent of statistical mechanics and fluid mechanics. 
A major difference is that swarming individuals are usually active (capable of consuming energy for propulsion), and social (capable of complex interactions with neighbors).
From a practical standpoint, understanding, predicting, and designing swarming behavior may find applications in topics as diverse as traffic optimization \cite{Aguilar2018}, crowd management \cite{Helbing2000}, video game design \cite{Boids}, and robotics \cite{Viragh2014}.
We envision it will become a central aspect of efficient management of floats of self-driving cars.

%Traditionally, the study of swarms has focused on understanding spontaneous symmetry breaking, i.e. how the swarm acquires a well-defined mean velocity $\langle \vec{v} \rangle$.
%This problem has seen two main approaches.
The study of swarming has traditionally encompassed two main approaches \cite{Ouellette2015}.
The first approach considers a discrete ensemble of agents, each with a velocity $\vec{V}_i(t)$, and investigates the ensemble properties of the system for different interaction rules between neighboring agents, and amount of noise \cite{Vicsek1995}.
The second approach, developed notably by Toner and Tu \cite{Toner2005}, is an extension of fluid mechanics, and the swarm is instead considered as a continuous medium, described by a velocity field $\vec{v}(\vec{r},t)$ and a density field $\rho(\vec{r},t)$.
General equations of motion can be written, with several adjustments compared to the usual Navier-Stokes equations of fluid mechanics, to account for the specifics of active systems.
Most notably, the incompressibility condition ($\nabla \cdot \vec{v} = 0$) usually does not hold.
Both approaches have different strengths and limitations, and their convenience depends on the system studied.
The hydrodynamics approach is better suited to large and dense crowds, where each individual's behavior is effectively lost among its peers.
Practically, it relies on techniques such as Particle Image Velocimetry or image registration to measure local displacements in complex images \cite{Silverberg2013,Bain2019}.
Conversely, agent-based models are adapted to low-density systems, where each individual can be tracked.

For some of the reasons mentioned above, sheep flocks have been the focus of several studies on collective behavior \cite{King2012,Ginelli2015,Strombom2014,Garcimartin2015}, although rarely from a hydrodynamics point of view.
In this paper, we analyze data from a publicly available aerial video of a large and \textit{dense} flock of sheep under stress from a shepherding dog.
The video is striking for the fluidity of flock movement.
We use simple video processing tools to analyze the data from a hydrodynamics perspective, by extracting 
%density and 
velocity \textit{fields}, as well as individual trajectories when resolvable. 
From them, we investigate propagation of information in the form of waves and correlations, and present simple observations about collective dynamics.

A few features are particularly interesting.
The flock is dense, with a quasi-stationary center of mass, and responds to perturbations from the shepherding dog. 
Therefore, we are looking at induced fluctuations about an equilibrium state in a jammed aggregate.
We find that while the flock tries to maintain a circular shape, significant motion happens at the edges, inducing long-range correlations spanning the entire flock.
% wave propagation

%response to perturbation -> ouellette
%dense
%no veolicty

%\section{Elements of sheep behavior and shepherding dog heuristics}

\section{Data and methods}

\begin{figure*}[h]
\includegraphics{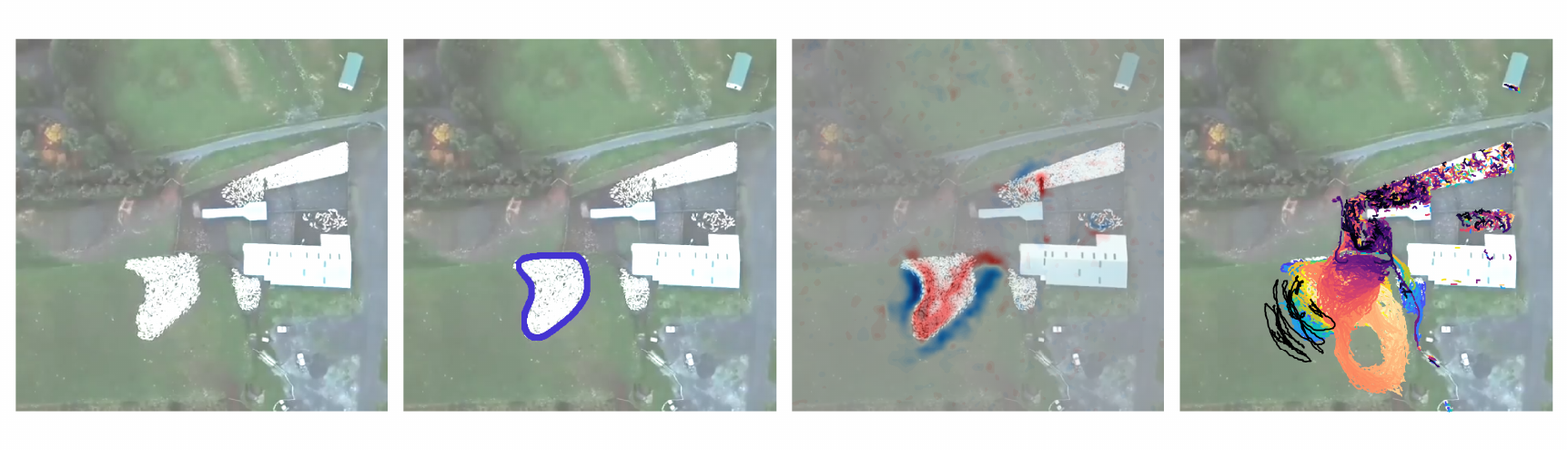}
\setlength{\unitlength}{1cm}
\put(-17,4) {\textbf{\textcolor{white}{(a)}}}
\put(-12.6,4) {\textbf{\textcolor{white}{(b)}}}
\put(-8.3,4) {\textbf{\textcolor{white}{(c)}}}
\put(-4,4) {\textbf{\textcolor{white}{(d)}}}
\put(-15,2.1) {\scriptsize \textsc{\textcolor{red}{SG}}}
\put(-17,1) {\scriptsize \textsc{\textcolor{red}{OF}}}
\put(-14.9,1.7) {\scriptsize \textsc{\textcolor{red}{SF}}}
\put(-15,2.8) {\scriptsize \textsc{\textcolor{red}{TF}}}

\caption{\label{fig1}
\textbf{Presentation of the movie and its analysis.}
(a) Frame of the movie at $t = 1$~s \cite{Movie}, and annotations (red text) corresponding to the description in Sec.~\ref{description}.
(b)	Outline (purple) of the flock of interest. See also Fig.~\ref{fig2}a.
(c) Superimposed heatmap of vorticity, calculated from image registration. See also Fig.~\ref{fig2}d and Movie~S3.
(d) Resolvable trajectories extracted throughout the movie: sheep trajectories in blue to yellow (Phase~1) to purple, indicating progression in time; dog trajectory in black.
}
\end{figure*}
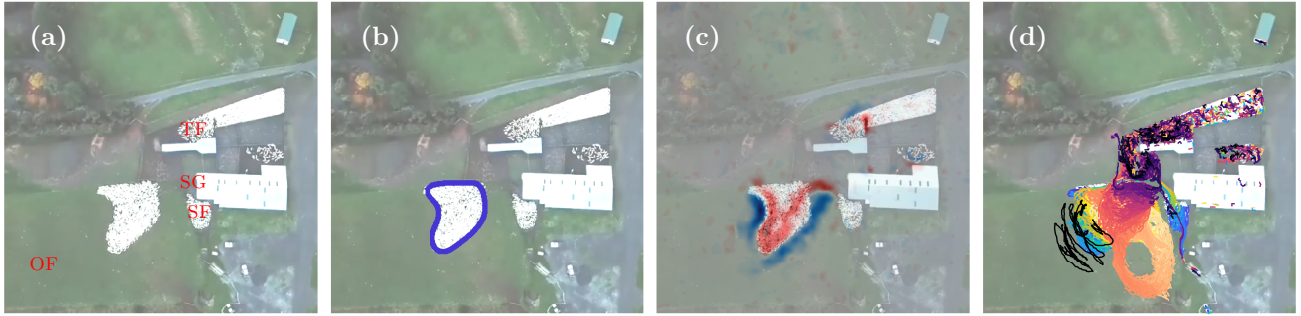

% do transparent
% change outline color

\subsection{Dataset}

The movie we analyze, but which did not record ourselves, is available on YouTube \cite{Movie}, and shows an aerial view from a drone over a sheep farm (Fig.~\ref{fig1}a).
Its duration is about 1~min, with a frame rate of 30~frames/s and a total of 1773 frames.
The field of view is 480$\times$480~pixels$^2$.
It is difficult to establish a correspondence between the time and space scales in the video with real scales.
The video appears accelerated, and cut at a few timestamps.
Because most of our analysis is qualitative, we identify the physical time $t$ with the video reference time, $t = t_\mathrm{ref}$.
In order to facilitate comparison with other studies, we do attempt to use roughly accurate space scales: measuring the length of the pick-up truck in the bottom-right corner at about 25~pixels, and comparing to a typical length for such a truck, 6~m, we can roughly establish the correspondence 1~m = 4~pixels.
The video shows different groups of sheep, some already parked in a fenced area.
We focus on the free flock at the center of the frame (Fig.~\ref{fig1}b).
In Section \ref{description}, we briefly describe the main events happening through the movie.

Using publicly available datasets such as YouTube videos, while untraditional, has proven a robust and inexpensive way to perform exploratory research on topics where experimental details are not crucial \cite{Silverberg2013,Yang2014,Yang2017}.

%\subsection{Fields of interest}
\subsection{Measurements of interest}
Following a fluid mechanics approach, we measure 
%density $\rho(\vec{r},t)$ and velocity $\vec{v}(\vec{r},t)$ fields, and calculate some time- and space-derivatives.
the velocity field $\vec{v}(\vec{r},t)$, and some of its space-derivatives.
To facilitate our analysis, we focus on the scalar fields that can be obtained.
Notably, we use divergence $\delta = \nabla \cdot \vec{v}$ as a measure of local expansion or contraction, and vertical vorticity $ \omega = \hat{k} \cdot \left( \nabla \times \vec{v} \right)$, which is a measure of both shear and curvature \cite{Bell1993}.
Other first order scalar fields can be calculated (see for instance \cite{Nave2019}), but we have found that divergence and vorticity are the most common and informative.
When possible, we also extract trajectories of individual sheep, and from them individual sheep $i$ velocities, $\vec{V}_i(t)$.

The density field $\rho(\vec{r},t)$ is also an important aspect of the dynamics, but its measurement is difficult due to the limited resolution of the video. 
We include a short discussion about density in the Supplemental Material.

%To facilitate further analysis, we will focus in the following mostly on the scalar fields that can be obtained.
%Looking at the Toner-Tu equations of motion, and using vector calculus identities, the $\vec{v}$-derived scalar quantities that enter the equation are the squared velocity $\vec{v}^2$, the divergence $\delta = \nabla \cdot \vec{v}$, and the (vertical) vorticity $ \omega = \hat{k} \cdot \left( \nabla \times \vec{v} \right)$.

%\subsection{Physical interpretation}
%The squared velocity $\vec{v}^2$ is simply a measure of how fast the flock moves locally.
%The divergence field $\delta$ is a measure 
%Vorticity measures \textit{local} rotation of a fluid element, and can be decomposed into two contributions: shear and curvature. 
%A large (absolute) vorticity means that different layers of the flock are moving at different velocities, but renormalized by the curvature of the local flow.
%WRONG In other words, wherever the vorticity is non-zero, two sheep moving initially side-by-side will not remain side-by-side, independently on the shape of their trajectories.
%The sign only indicates the direction of rotation, clockwise or counterclockwise.

\subsection{Methods}

Image analysis and data analysis employed custom code written in MATLAB.
%The density field was simply inferred from measurements of pixel brightness (sheep are white) combined with a local averaging procedure (Gaussian kernel).
The calculation of the velocity field was based on image registration, a technique that calculates \textit{local} deformations (translation and rotation) between two images. 
The MATLAB built-in function used was \verb"imregdemons".
Tracking of individual sheep, where resolvable, relied on finding local maxima in brightness and connecting them between frames into trajectories, based on the code provided in Ref.~\cite{Crocker1996}.
Finally, the tracking of the shepherding dog was done manually.
Correction for various artifacts involved moderate spatial and/or temporal averaging.

Further details about processing methods are provided in the Supplemental Material, along with assessment of reliability. 
The MATLAB code is also available from the authors' GitHub page \cite{GitHub}.

\section{Interpretative description of events and behavior in the video}
\label{description}
We briefly describe here the main events happening in the video, and relate them to some heuristics of sheep-dog behavior provided by shepherds.
\textit{Nota bene}: shepherds often use the term ``pressure'' to refer to the influence of a specific stimulus (typically the dog) on the flock that tends to push the flock away; therefore, we use this term in this section as well.
However, because the dog's influence on the flock does not rigorously meet the physics definition of pressure (an isotropic stress tensor), we refrain from using this term in the quantitative discussion that follows. 

When the video starts, the side gate (Fig.~\ref{fig1}a, SG) is open, and sheep are coming into the main flock, wrapping around the bottom edge in a clockwise direction, adding pressure and initiating a swirling response (Fig.~\ref{fig2}e). 
At $t = 1$~s, the side gate is closed, and the main flock continues to swirl due to two pressures: 1) the incoming sheep, 2) the dog covering the open field (OF) on the left side of the frame.
Sheep tend to move away from pressure and follow the movement of other sheep, using visual clues.
The side flock (SF) funnels through from the side pen to the top pen to join the top flock (TF). 
At $t = 10$~s, the main flock bunches up near the top gate, in response to seeing the side herd flow into the top pen.
At $t = 13$~s, the opening of the top gate causes a disturbance in these sheep at the top edge of the main flock, sending them away from the gate and downward along the side fence to the bottom fence. 
At $t = 21$~s, the top gate is fully open and the sheep begin to flow through, going towards ``negative pressures'', \textit{i.e.} empty space. 
This creates a collective alignment of all sheep towards the gate.
At $t = 23$~s, there is a complete break between sheep in the main flock and sheep along the bottom fence. 
There is a 2~s time lag before the sheep at the bottom start to re-join the main flock and follow their motion. 
At $t = 24$~s, there is another external disturbance: a break in the fence.
Four sheep escape, but the fence is re-secured between 25~s and 26~s. 
Although this causes a small disturbance, the main flock starts to go through the top gate again at $t = 27$~s. 
By $t = 28$~s, the bottom group has completely re-joined the main flock and the flock continues to funnel through the top gate.
The dog increases pressure from $t = 30$~s to $t = 33$~s to get sheep through the gate faster.
The rest of the video only shows the dog and the shepherd trying to bring some small dispersed groups into the pen.
Although the shepherd is present throughout the video, its role is most likely to communicate commands to the dog, which is the proxy that interacts directly with the flock.

%That flock will be rounded up by a dog, and finally pushed into the enclosure.
%There are two main phases in the flock motion.
To facilitate our analysis, we differentiate two main phases.
Phase~1 goes from $t = 0$~s to $t = 24$~s.
The flock is ``collected" by the dog and fluctuates around a circular shape, where instabilities develop leading to pinching and breaking of the flock into two groups.
Phase~2 goes from $t = 28$~s to $t = 36$~s, and the flock progressively flows into the enclosed area.

\section{Results}

\subsection{Phase 1: dynamics of a free flock under stress from a shepherding dog.}

%In this phase (Figure~\ref{fig2}, $t \leq $ 24~s), the flock has no net motion, and fluctuations develop around an equilibrium state, triggered by the dog.

\begin{figure*}[h]
\includegraphics{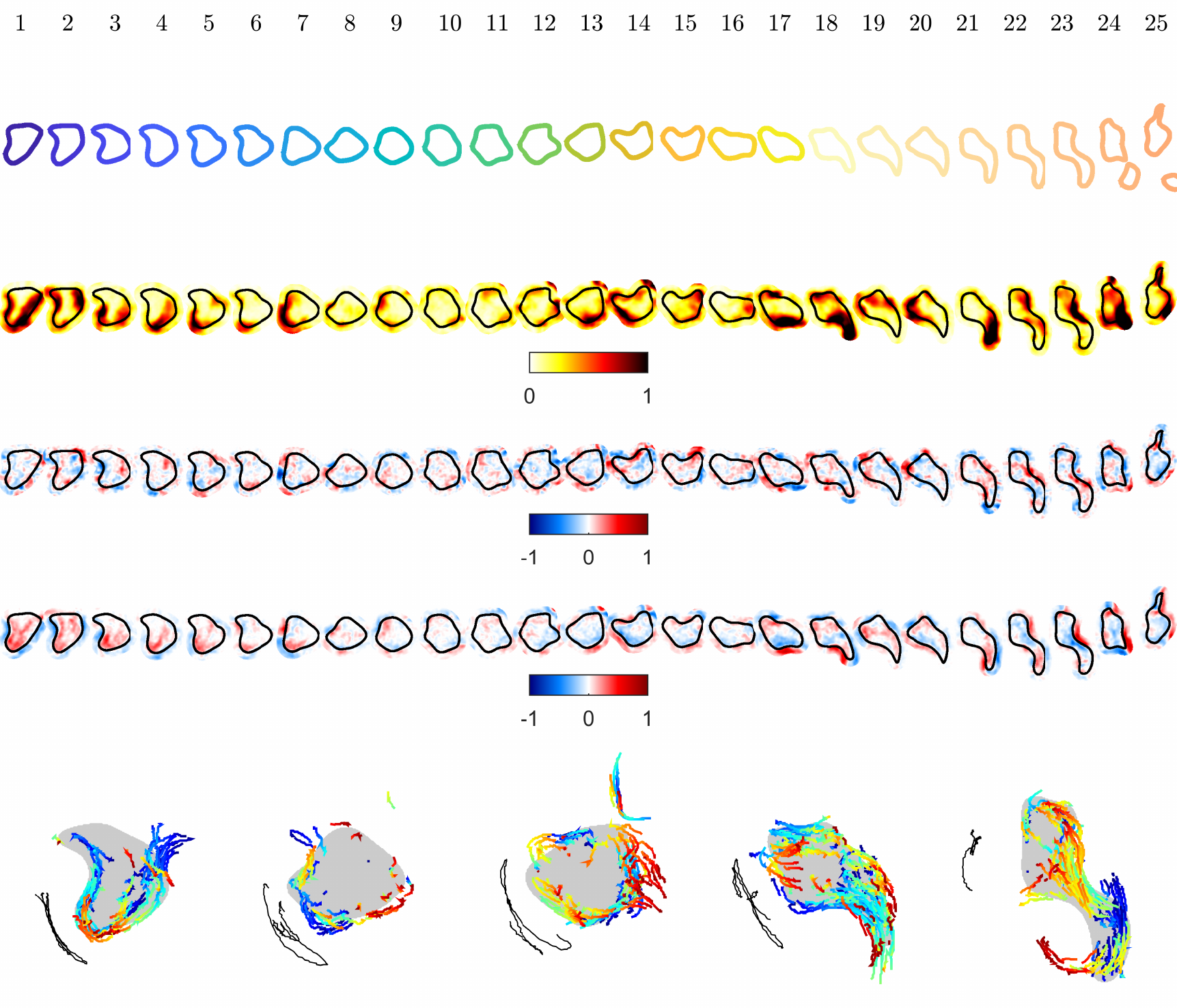}
\setlength{\unitlength}{1cm}
\put(-17.4,13.5) {\textbf{(a)} - outlines}
\put(-17.4,11) {\textbf{(b)} - velocity $\vert v \vert$}
\put(-17.4,8.6) {\textbf{(c)} - divergence $\nabla \cdot \vec{v}$}
\put(-17.4,6.1) {\textbf{(d)} - vorticity $\vert \nabla \times \vec{v} \vert$}
\put(-17.4,3.5) {\textbf{(e)} - trajectories}
\caption{\label{fig2}
\textbf{Free flock geometry and hydrodynamic metrics at each second ($t = 1 \dots 25$~s, see first row).}
(a)~Free flock outlines, emphasizing a wavy pattern with time-dependent undulations.
(b)~Local velocity (in magnitude), inferred from velocity field measurements, and showing a concentration of motion at the edges. Scale bar in arbitrary units.
(c)~Divergence fields (calculated as $\nabla \cdot \vec{v}(\vec{r},t)$) often showing a concentration of divergence on areas of high curvature. Scale bar in arbitrary units.
(d)~Vorticity fields (calculated as $\nabla \times \vec{v}(\vec{r},t)$, in magnitude), localized mostly along the edges. Scale bar in arbitrary units.
(e)~Individual trajectories over 5~s intervals, for resolvable sheep (colored trajectories, showing progression over time from blue to red). Dog trajectory over similar intervals is plotted in black.
}
\end{figure*}

In Phase~1, the flock is ``collected" by the dog.
While the flock shows no net motion, sheep are locally moving and exchanging positions (Fig.~\ref{fig2}e), so that fluctuations develop around an equilibrium state.
%Eventually, these fluctuations diverge, leading to the breaking of the flock into two groups. 
Eventually, these fluctuations amplify, leading to the breaking of the flock into two groups.

%\begin{figure*}[h]
%\includegraphics{fig2.pdf}
%\setlength{\unitlength}{1cm}
%\put(-17.4,13.1) {\textbf{$\partial S$}}
%\put(-17.4,11.3) {\textbf{$\rho$}}
%\put(-17.4,9.3) {\textbf{$v$}}
%\put(-17.4,7.5) {\textbf{$\delta = \nabla \cdot \vec{v}$}}
%\put(-17.4,5.6) {\textbf{$\omega = \vert \nabla \times \vec{v} \vert$}}
%\put(-17.4,4) {\textbf{$\varphi$}}
%\caption{\label{fig2}
%Free flock geometry and hydrodynamic metrics at each second:
%outlines ($\partial S$),
%density ($\rho$),
%velocity ($v$),
%vertical vorticity ($\omega$),
%divergence ($\delta$).
%Individual trajectories over 5~s intervals ($\varphi$) are also displayed, when possible.
%}
%\end{figure*}

%\subsubsection{Under stress, the flock's preferred shape is circular, but the equilibrium is unstable.}
\subsubsection{Under stress, the flock's preferred shape is circular.}
\label{CircularEquilibrium}
Fig.~\ref{fig2}a shows that the shape of the flock evolves from undulating ($t = 1 \ldots 7$~s), to circular ($t = 8 \ldots 10$~s), to undulating again, and eventually fragmentation ($t = 24$~s).
The circular phase is characterized by a minimum in kinetic energy (Fig.~\ref{fig2}b). %and most homogeneous density (Fig.~\ref{fig2}$\rho$).
The concentration of divergence on certain pointed areas of the edge (Fig.~\ref{fig2}c, $t = 2,15,20$~s) suggests that the flock tends to erase protrusions.
The circular equilibrium, however, 
%is unstable, and 
quickly collapses when a gate opens in the top-right corner at $t = 13$~s.
%the dog resumes its pressure.
From then, fluctuations amplify, and eventually trigger the collapse of the flock's cohesiveness.

%These observations may be connected

% comparison shape "energy"
% divergence concentrated at the pointy edges
% add dog

%%%\subsubsection{Sheep density is non-uniform, higher near the dog, and lower at the edges.}
%%%
%%%A flock is generally not incompressible, as the distance between individuals can vary from contact to the full range of sensory interactions.
%%%Indeed, Figure~\ref{fig2}$\rho$ shows that the free flock, while maintaining strong cohesiveness, also exhibits significant variations in local density.
%%%The density is lower at the edges, as easily seen in Movie~S1 and from the trajectories that are resolvable (Fig.~\ref{fig2}$\varphi$).
%%%The density tends to be higher in areas directly facing the dog, as sheep under increased stress tend to pack up more stiffly \cite{Ginelli2015}.
%%%These disparities in density induce differences in how much sheep can move locally, and how much visual clues they can perceive.

\subsubsection{Sheep motion is concentrated at the edges of the flock.}
Fig.~\ref{fig2}b shows that areas of high velocity are distributed along the flock's edges.
The vorticity maps in Fig.~\ref{fig2}d further confirm that these edges are areas of high shear, so that while sheep at the edge are moving, sheep deeper inside the flock are not.
These observations are consistent 
%with density distributions, as 
with the fact that sheep inside the flock have little freedom of motion, and lack the visual clues to decide whether, and where, to move.
Interestingly, high velocities are only localized on a fraction of the edge, and move along it (Fig.~\ref{fig2}; e.g., $t = 4 \ldots 7$~s).
These areas often correspond to excitation from the dog (Fig.~\ref{fig2}b,d,e, $t = 7,14$~s  ), but not always.
Sometimes, the sheep's persistence to follow others or a specific goal creates large displacements as well ($t = 18,21,23$~s).

\subsubsection{The flock enables the propagation of edge waves.}
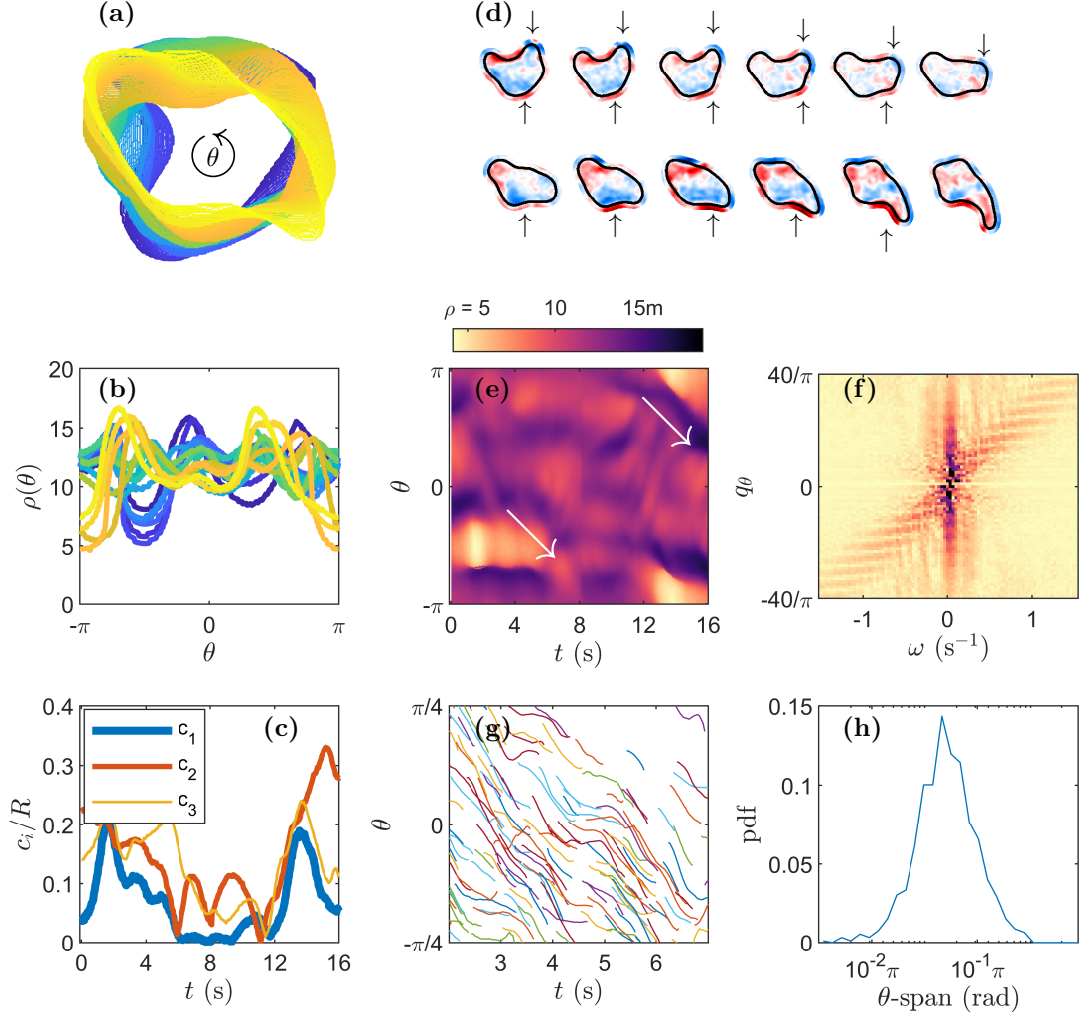
\begin{figure*}[h!]
\includegraphics{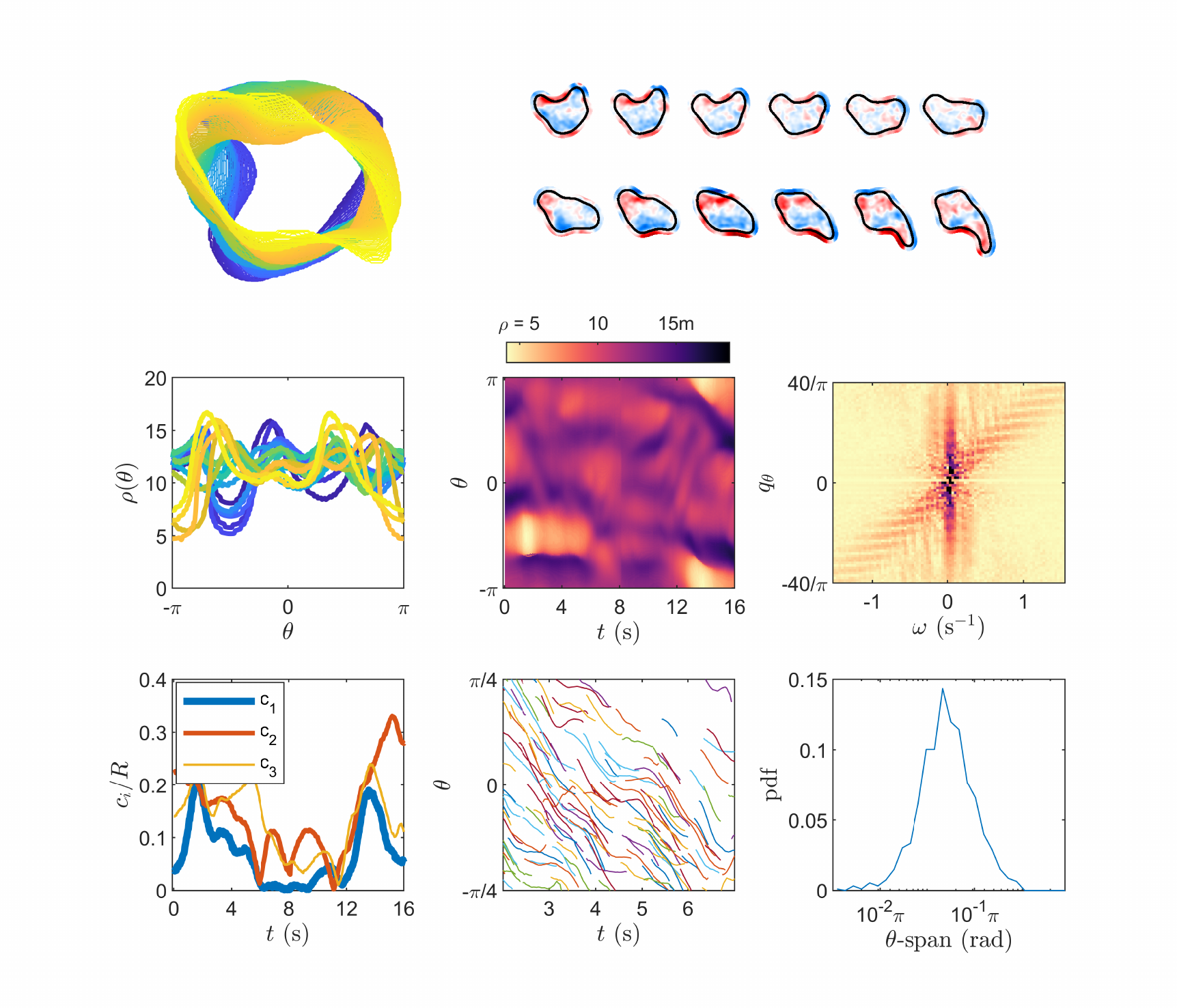}
\setlength{\unitlength}{1cm}
\put(-14.7,14) {\textbf{\textcolor{black}{(a)}}}
\put(-14.7,9) {\textbf{\textcolor{black}{(b)}}}
\put(-12.5,4.5) {\textbf{\textcolor{black}{(c)}}}
\put(-9.7,14) {\textbf{\textcolor{black}{(d)}}}
\put(-9.7,9) {\textbf{\textcolor{black}{(e)}}}
\put(-4.8,9) {\textbf{\textcolor{black}{(f)}}}
\put(-9.7,4.5) {\textbf{\textcolor{black}{(g)}}}
\put(-4.8,4.5) {\textbf{\textcolor{black}{(h)}}}
\put(-9.1,11.2) {$\mathbf{\uparrow}$}
\put(-7.9,11.2) {$\mathbf{\uparrow}$}
\put(-6.6,11.2) {$\mathbf{\uparrow}$}
\put(-5.5,11.2) {$\mathbf{\uparrow}$}
\put(-4.3,11.0) {$\mathbf{\uparrow}$}
\put(-9.1,12.7) {$\mathbf{\uparrow}$}
\put(-7.9,12.7) {$\mathbf{\uparrow}$}
\put(-6.6,12.7) {$\mathbf{\uparrow}$}
\put(-5.4,12.7) {$\mathbf{\uparrow}$}
\put(-4.3,12.7) {$\mathbf{\uparrow}$}
\put(-9.0,13.9) {$\mathbf{\downarrow}$}
\put(-7.8,13.9) {$\mathbf{\downarrow}$}
\put(-6.6,13.9) {$\mathbf{\downarrow}$}
\put(-5.4,13.8) {$\mathbf{\downarrow}$}
\put(-4.2,13.7) {$\mathbf{\downarrow}$}
\put(-3.0,13.6) {$\mathbf{\downarrow}$}
\put(-7.5,8.5)  {\huge{\textcolor{white}{$\searrow$}}}
\put(-9.3,7)  {\huge{\textcolor{white}{$\searrow$}}}
\put(-13.5,12) {\Huge{$\circlearrowleft$}}
\put(-13.25,12.08) {{$\theta$}}
\caption{\label{figWave}
\textbf{Wave propagation along the flock's edge in Phase~1, $\mathbf{t \leq 16}$~s.}
(a) Superimposed flock outlines over time, from blue to yellow, showing evolving undulations.
(b)	Radial decomposition of flock shapes, from $(x,y)$ to $\rho(\theta)$. Different radial profiles are shown at different times, corresponding to the shapes in (a).
(c) Fourier series decomposition of $\rho(\theta)$ over time. 
The first three Fourier coefficients $c_i(t)$, normalized by the zeroth component $R$, are shown as a function of time.
A low amplitude in the coefficients, for example for $t = 6 \dots 10$~s, indicates a more circular shape. A high amplitude indicates a wavy shape.
(d) Wave propagation evidenced in two different sequences of the vorticity field.
Black arrows indicate high vorticity moving along the edge.
(e) Kymograph of the radial profiles $\rho(\theta)$. 
Dark colors indicate peaks in $\rho(\theta)$, and light colors indicate dips (see color bar).
White arrows guide the eyes along the motion of some peaks and dips in the $(t,\theta)$-plane.
(f) Power spectrum of $\rho(\theta,t)$, normalized at every wave vector.
The high values on the diagonal indicate an approximately linear dispersion relation.
(g) Representative edge trajectories over a small window of space and time, showing a succession of short trajectories, rather than a few long trajectories.
(h) Histogram of the angle span of resolvable trajectories, about 1000 of them. 
Most trajectories only cover about 0.1~$\pi$~rad, which is small compared to the typical angular propagation of waves in (e), around $\pi$~rad.
}
\end{figure*}

The wavy aspect of the flock shape is readily visible in Movie~S1 and in Fig.~\ref{fig2}a.
In order to characterize this pattern, we convert the parametric coordinates of the flock outlines $(x,y)$ (Fig.~\ref{figWave}a) into radial profiles $\rho(\theta)$ (Fig.~\ref{figWave}b), for $\theta$ in $\left(-\pi ; \pi \right]$ and at every $t \leq 16$~s \cite{Pecreaux2004}.
The Fourier series decomposition of $\rho(\theta)$ in Fig.~\ref{figWave}c confirms the observations of \ref{CircularEquilibrium}: the decrease in the coefficients' amplitude between $t = 0$~s and $t = 6$~s corresponds to the flock becoming more circular, while the sharp increase in their amplitude after $t = 12$~s correspond to the large undulations observed in the flock's shape.
%, showing a``calming" of the flock's undulations through $t = 6$~s, followed by amplifying fluctuations after $t = 10$~s .

The question now is whether these shape fluctuations propagate over space and time like waves do.
It seems to be the case, from careful examination of the Supplemental Movies.
For example, Fig.~\ref{figWave}d presents two different sequences where the vorticity field is clearly seen propagating along the edge.
For a rigorous analysis, we present the kymograph of the radial profiles in Fig.~\ref{figWave}e.
It shows the peaks and dips in $\rho(\theta,t)$ moving along $\theta$ over time.
In Fig.~\ref{figWave}f, the corresponding power spectrum (two-dimensional Fourier transform) is peaked along the diagonal in the plane defined by frequency $\omega$ and wave vector $q_\theta$, which indicates a linear dispersion relation at small $q_\theta$ \cite{Bain2019}.
This is strong evidence that the fluctuations in the flock's shape indeed propagate along the edge.

In order for these fluctuations to qualify as waves, they need to be carried \textit{through} the sheep, but not \textit{by} the sheep. 
In other words, they must not correspond to a few sheep running continuously around the flock.
We verify this hypothesis by looking at individual sheep trajectories in Fig.~\ref{figWave}g.
They reveal that for a $\rho(\theta)$ peak moving a distance about $\pi/2$ over about 6~s, there is a large number of sheep covering only a small fraction of that distance.
Individual sheep move a small distance then stop, while other sheep in the vicinity start and continue further.
These trajectories show a relay-run type of propagation, as further evidenced by the distribution of trajectories angular-spans (Fig.~\ref{figWave}h), showing that most trajectories only cover a small amount of the total arc-distance covered by deformations.
% explain why trajectories disappear

In conclusion, shape undulations propagate in space and time along the edge of the flock, and have the microscopic properties of waves, whereby no large-scale transport of matter happens (this is only true until the flock starts fragmenting, up to $t = 16$~s).

\subsubsection{Edge motion is instantaneously correlated across the entire flock.}

\begin{figure}[h!]
\includegraphics{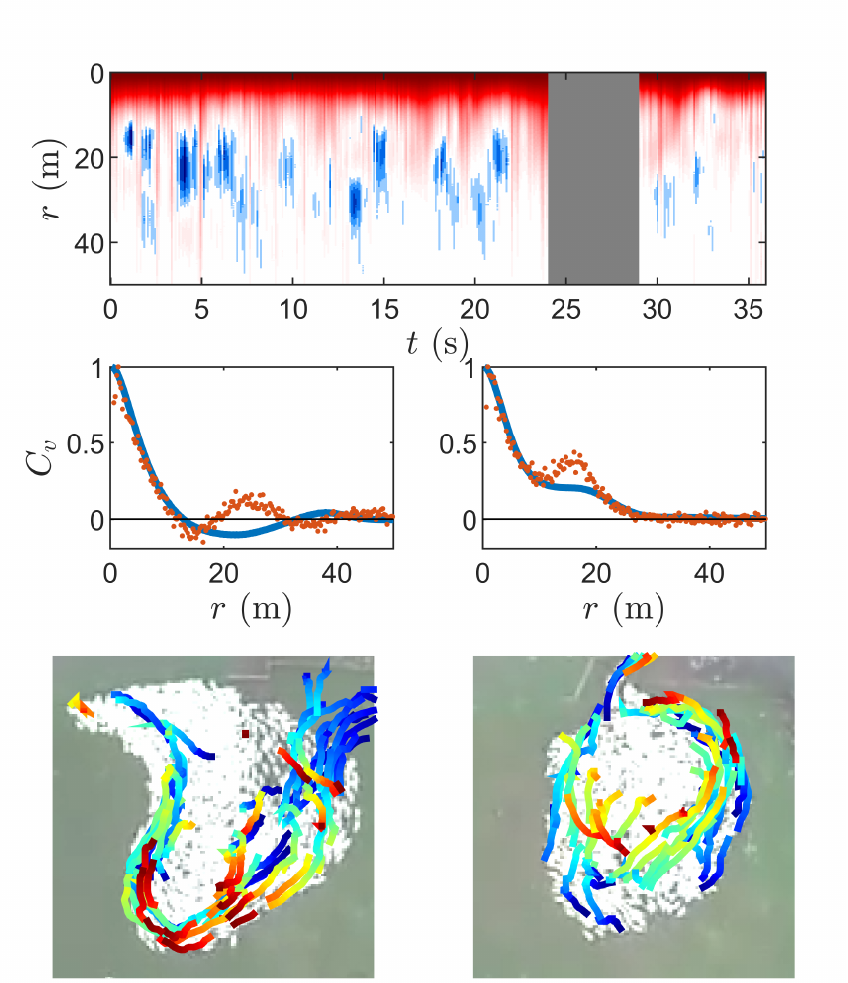}
\setlength{\unitlength}{1cm}
\put(-8.3,9.5) {\textbf{\textcolor{black}{(a)}}}
\put(-8.3,6.3) {\textbf{\textcolor{black}{(b)}}}
\put(-1,6.3) {\textbf{\textcolor{black}{(c)}}}
\put(-8.3,3.4) {\textbf{\textcolor{black}{(d)}}}
\put(-1,3.4) {\textbf{\textcolor{black}{(e)}}}
%\put(-4,3.8) {\textbf{\textcolor{black}{(f)}}}
\caption{\label{figCorr}
\textbf{Sheep velocity correlations in Phase~1 and Phase~2.}
(a) Correlations in the velocity field $\langle [\vec{v} \cdot \vec{v}](r) \rangle$ as a function of distance $r$ and at various times (Phase~1 up to 24~s, Phase~2 after 29~s). 
Red indicates positive correlations, and blue negative correlations.
(b,c) Normalized correlations from velocity fields (blue solid line) and individual trajectories (orange dotted) as a function of distance $r$ at $t = 4$~s (b)	and $t = 30$~s, showing respectively an anticorrelation or a correlation at $r \simeq 20$~m, roughly the size of the flock. 
(d,e) Representative sheep trajectories over 4~s time intervals (blue to red), in (d) Phase~1 ($t = 4$~s), and (e) Phase~2 ($t = 30$~s).
}
\end{figure}

While waves propagate at \textit{finite} speed, we also observe \textit{instantaneous} velocity correlations over distances comparable to the flock's diameter.
In Fig.~\ref{figCorr}, we calculate velocity correlations within single frames, both from the velocity field $C_v(r) = \langle \vec{v}(\vec{x}) \cdot \vec{v}(\vec{x}+\vec{r}) \rangle_{\vec{x}}$ and from individual trajectories $C_V(r_{ij}) = \langle \vec{V}_i \cdot \vec{V}_j\rangle_{ij} $ \cite{Ni2015}.
The results from these two metrics are qualitatively similar (Fig.~\ref{figCorr}b).
They show that velocities, naturally correlated at short distances, are also anticorrelated at separations around 20~m, which corresponds to the distance between two opposite edges of the flock (Fig.~\ref{figCorr}a, $t \leq$~24~s).
This is a signature of a general swirling dynamics of the flock \cite{Silverberg2013}, which is further confirmed looking at Movie~S1 and at representative trajectories in Fig.~\ref{figCorr}d.
The apparent intermittent nature of the swirling (Fig.~\ref{figCorr}a) may indicate that it occurs when more than one wave propagate at the same time, due to slow attenuation and continued excitation from the dog (more details in \ref{sheepDog}).

The flock motion in Phase~2 shows a different dynamics, with reversed correlations, which will be examined below.
%%% correlation lag at the end

%Figure~\ref{figCorr} presents \textit{instantaneous} velocity correlations of the flock in Phase~1 and Phase~2.
%The motion of sheep on the edge of the flock appears to induce long-range correlations.
%In Phase~1, when the flock is overall swirling, anticorrelations develop over distances much larger than typical sheep-sheep alignment, and equal to the sheep diameter (Fig.~\ref{figCorr}a).
%This is apparent in both the velocity field and with individual trajectories (Fig.~\ref{figCorr}b).
%This is consistent with what trajectories reveal about sheep motion around the edge, as visible in Fig.~\ref{figCorr}d, showing that sheep generally move along the edge in the same direction.
%Importantly, because these correlations are instantaneous, this means that the flock is immediately correlated.

%This time, velocities show a correlation peak from one edge of the flock to the next (Fig.~\ref{figCorr}a,c).
%In both instances, information has percolated through the flock.
% correlations

\subsection{Phase 2: spontaneous flow of the flock through a small aperture.}

\begin{figure*}[h!]
\includegraphics{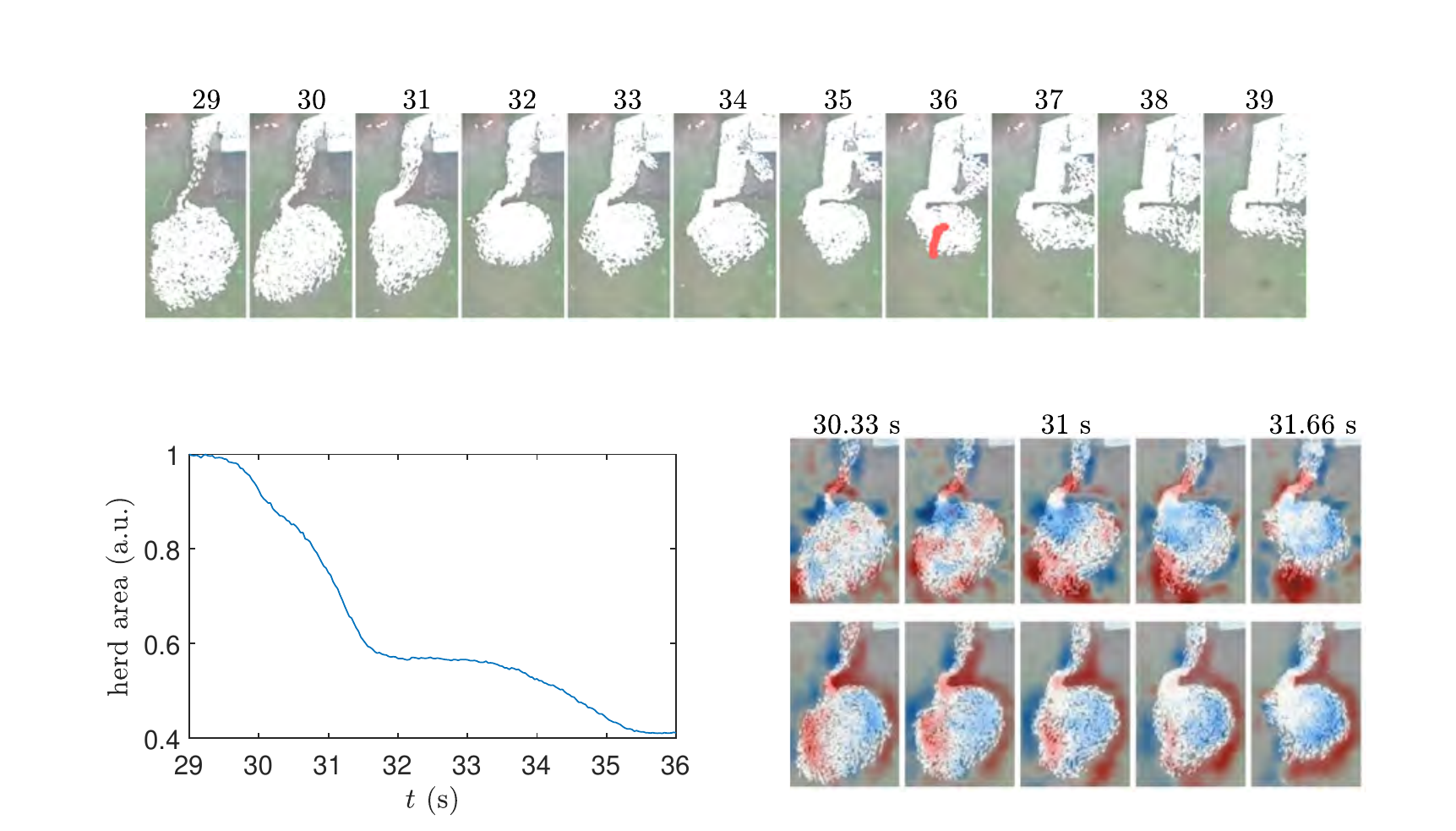}
\setlength{\unitlength}{1cm}
%\put(-15,14) {\textbf{\textcolor{black}{(a)}}}
%\put(-15,9.7) {\textbf{\textcolor{black}{(b)}}}
%\put(-8.8,9.7) {\textbf{\textcolor{black}{(c)}}}
%\put(-15,5.1) {\textbf{\textcolor{black}{(d)}}}
%\put(-8.8,5.1) {\textbf{\textcolor{black}{(e)}}}
\put(-16,9) {\textbf{\textcolor{black}{(a)}}}
\put(-16,5) {\textbf{\textcolor{black}{(b)}}}
\put(-8.6,4.5) {\textbf{\textcolor{black}{(c)}}}
\put(-8.6,2.3) {\textbf{\textcolor{black}{(d)}}}
\put(-15,7.5) {\large{$\nwarrow$}}
\put(-7.5,3.5) {\large{$\nwarrow$}}
\put(-6,3.4) {\large{$\nwarrow$}}
\put(-4.5,3.3) {\large{$\nwarrow$}}
\put(-3,3.2) {\large{$\nwarrow$}}
\put(-1.5,3.1) {\large{$\nwarrow$}}
\caption{\label{figDischarge}
\textbf{Flow of the flock into the pen.}
(a) Image sequence showing the flock flowing into the pen \cite{Movie}, at different times between $t = 29$~s and $t = 39~$s.
The black arrow indicates the cohesive jet of sheep.
The trajectory of the center of mass (for $t \geq 29$~s) is overlaid in red on the frame at 36~s.
(b)	Area of the flock (arbitrary units) outside the pen, as a function of time. The flock flows quickly at the beginning, then stops, then resumes at a slower pace.
(c) Divergence field around 31~s, showing notably the fast contraction of the flock and the associated propagation of a high divergence patch (blue, annotated with black arrow) from the gate to the edge.
(d) Vorticity field around 31~s, showing a strong current of sheep on the edge of the flock.
}
\end{figure*}

In Phase~2, the flock spontaneously flows through a gate to reach an enclosed area (Fig.~\ref{figDischarge}).
We can assume that the sheep collectively know where to go, which results in motion that is less confused than in Phase~1, but the dog also induces an additional 
%pressure
force on the flock.
There is a net, slow motion of the center of mass, as the flock moves from one area into a closed container.
When the pen is full, the flow stops.

%%% filling???

\subsubsection{Flock motion is ordered and correlated.}

It is apparent from Movie~S1 and Fig.~\ref{figDischarge}a that the flock flows slowly and continuously through the gate, with no fluctuations in shape.
Positive correlations in sheep velocities extend across the flock (Fig.~\ref{figCorr}c), as sheep on the inside push through the gate, and sheep on the edge run around it to enter the pen (Fig.~\ref{figCorr}e).
It can be safely assumed that in this phase, due to the flock configuration and its position relative to the farm, individual sheep know where the gate is and that they have to enter through it, even with limited visual clues.
This leads to a spontaneous and polarized dynamics.
However, the dog can also be seen ``pushing" the flock to accelerate the flow, for $t = 30 \ldots 33$~s.
This results in an acceleration of the flow into the pen (fast decrease in the flock area around $t = 31$~s in Fig.~\ref{figDischarge}b), as well as a fast contraction of the flock on itself, visible in Fig.~\ref{figDischarge}a and in the propagation of the divergence field inside the flock in Fig.~\ref{figDischarge}c.
The flock, of course, is only compressible to a certain extent, and when the pen becomes full at $t = 31.5$~s, the flow stops (constant flock area in Fig.~\ref{figDischarge}b), and only resumes when sheep start flowing into a secondary pen (Fig.~\ref{figDischarge}a,b $t \geq 33$~s).

\subsubsection{Strong cohesion in the flock is maintained during the flow.}

Interestingly, the flock remains in a strongly cohesive state as it flows into the pen, as seen in the time sequence in Fig.~\ref{figDischarge}a.
Sheep acceleration leads to larger average distance in the jet, visible notably at $t = 29, 30$~s, but even then sheep remain aligned and do not fill the pen uniformly.
This is further evidence that sheep are strongly attracted to each other, and follow one another closely, at least in the behavioral state observed here.

\subsubsection{Edge flow leads to mass redistribution, but circular shape persists.}

Aside from the flock becoming smaller as it fills the pen, its center of mass also shifts towards the right side of the gate (Fig.~\ref{figDischarge}a and red trajectory at $t = 36$~s).
The reason is that the strong current along the flock (Fig.~\ref{figDischarge}d) brings the sheep close to the gate, but due to the specific geometry, sheep have an easier access on the left side compared to the right side.
Therefore, while the left side becomes depleted of sheep, an accumulation occurs on the right side, bringing the center of mass in that direction.
The circular shape of the flock persists nonetheless, further confirming it is the preferred shape.

\subsection{Interactions between individual trajectories}
\label{sheepDog}

We now turn to individual trajectories.
By looking at spatial and temporal correlations, we attempt to reveal some of the microscopic mechanisms at the origin of the flock's large-scale dynamics.
Due to resolution limitations, individual trajectories are only detectable in areas of lower densities, typically on the edges.
The resolution on displacements is 1~pixel ($\sim$0.25~m), and hence the minimum resolvable distance between two sheep is 2~pixels.

\subsubsection{Sheep-dog velocity correlations.}

\begin{figure*}
\includegraphics{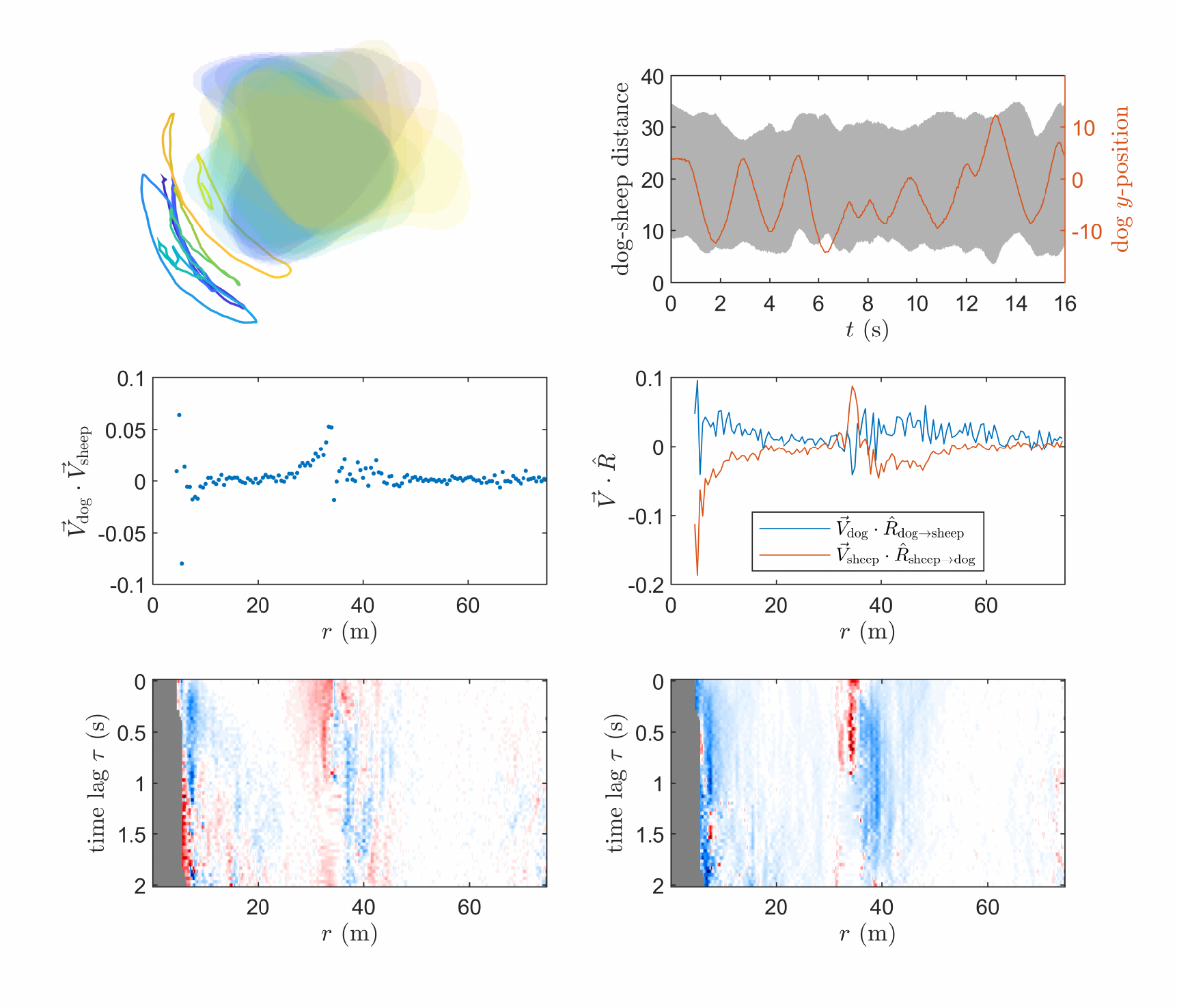}
\setlength{\unitlength}{1cm}
\put(-14.5,13.3) {\textbf{\textcolor{black}{(a)}}}
\put(-6.9,13.3) {\textbf{\textcolor{black}{(b)}}}
\put(-14.5,8.8) {\textbf{\textcolor{black}{(c)}}}
\put(-6.9,8.8) {\textbf{\textcolor{black}{(d)}}}
\put(-14.5,4.3) {\textbf{\textcolor{black}{(e)}}}
\put(-6.9,4.3) {\textbf{\textcolor{black}{(f)}}}
\caption{\label{figSheepDog}
\textbf{Interactions between sheep and the shepherding dog in Phase~1.}
(a) Dog trajectory over time (blue to yellow, same times as previous figures), and corresponding flock shape and position (transparent colored patches).
(b)	Left axis: distance between the dog ($y = 0$) and the flock (grey area). 
The dog maintains a constant distance (between 5~m and 10~m) with the closest edge of the flock (lower grey boundary).
Right axis: $y$-position of the dog in the frame showing back and forth motion with a period of about 2~s.
(c) Instantaneous velocity correlations between the dog and the sheep.
(d) Dog velocity (blue) projected on direction with sheep, and sheep velocity (orange) projected on direction with dog.
(e) Dog-sheep velocity correlations as a function of distance~$r$ and lag-time~$\tau$ (see text).
(f) Sheep velocities at $t$ projected onto their direction relative to the dog at $t-\tau$.
}
\end{figure*}

We focus here on dog-sheep interactions in Phase~1, when the dog is most active.
The dog maintains a constant distance from the flock (Fig.~\ref{figSheepDog}a,b), about 10~m away from the closest edge.
This is in conformity with dog shepherding heuristics, which typically refers to the constant space between the dog and the flock as the ``bubble''.
The dog covers by moving back and forth along the edge of the flock with a period of about 2~s (video time, see Fig.~\ref{figSheepDog}b).

We investigate how dog and sheep velocities are correlated as a function of distance, and how these correlations propagate over time.
We look at averaged correlations
\begin{equation}
\mathcal{C}(r,\tau) = \langle \vec{V}_\mathrm{dog}(t) \cdot \vec{V}_i(t-\tau,r) \rangle_{i,t},
\end{equation}
where $r$ is the dog-sheep distance, and $\tau$ the propagation (lag) time.
Negative correlations at short distances indicate that sheep run in the opposite direction of the dog (Fig.~\ref{figSheepDog}c).
%%% Interesting -- more to stay?
Interestingly, on the far edge ($r \simeq$ 30~m), these velocities are positively correlated.
This is likely a consequence of the overall swirling of the flock, which creates opposite velocities at opposite edges.
The short-range correlations become positive at lag times $\tau \geq$~1~s (Fig.~\ref{figSheepDog}), indicating that sheep's running direction is determined by the dog's direction, but does not change when the dog turns around.
% time propagation

\subsubsection{Sheep-dog orientations.}

To characterize how the dog interacts with sheep, we look at correlations $\phi_\mathrm{dog}$ between the dog's velocity at a time~$t$ and its position relative to a certain sheep~$i$ at a distance~$r$ at a prior time~$t-\tau$: $\hat{R}_{\mathrm{ds_i}} = (\vec{R}_i - \vec{R}_\mathrm{dog})/r$ \cite{Puckett2015}. 
Symmetrically, we look at how sheep react to the dog's displacements through sheep velocities projected onto the sheep $\rightarrow$ dog direction, $\hat{R}_{\mathrm{s_i d}}$.
Formally, we define: 
\begin{equation}
\phi_\mathrm{dog}(r,\tau) = \langle \vec{V}_\mathrm{dog}(t) \cdot \hat{R}_{\mathrm{ds_i}}(t-\tau,r) \rangle_{i,t},
\end{equation}
and
\begin{equation}
\phi_\mathrm{sheep}(r,\tau) = \langle \vec{V}_i(t) \cdot \hat{R}_\mathrm{s_id}(t-\tau,r) \rangle_{i,t}.
\end{equation}
For instantaneous correlations ($\tau = 0$), we find positive (attractive) correlations for the dog, and negative (repulsive) correlations for the sheep, at short distances (Fig.~\ref{figSheepDog}d).
Unsurprisingly, the dog tends to move towards the sheep, and the sheep away from the dog.
These correlations are reversed at the far edge ($r \simeq$ 30~m), due to the flock's swirling and cohesion.
The persistence of the negative peak in $\phi_\mathrm{sheep}(r,\tau)$ over $\tau$ likely indicates that the dog pushes against the flock consistently.

%\subsubsection{Sheep-sheep interactions.}
%Unlike swarms of flying insects, which are not dense (gas-like) and hence allow for fairly unconstrained motion and long-range propagation of information \cite{Cavagna2017}, the sheep flock is almost jammed, and visual clues are limited to the immediate vicinity of a sheep.
%Therefore, we focus on the instantaneous, short-range interactions, because they are the only ones directly related to a sheep's decision scheme.
%Because these individual interactions rules should be less sensitive to the kind of flock behavior, we consider trajectories extracted throughout the entire movie.

\section{Discussion}
%Ni Ouellette 2015: "correlation, rather than order is the hallmark of emergent collective behavior"..."important role of enviromental conditions on collective behavior"..."system with strong correlations show enhanced responses to external stimuli

We now discuss possible implications of the dynamical features reported above.
%It ought to be stressed, however, that the conclusions on this single flock cannot be fully generalized without additional observations and experiments.

\subsection{Strong correlations are a signature of collective behavior}

Collective behavior is often perceived as the property of a large number of individuals to follow each other towards a certain location.
This is why a number of theoretical models have focused on understanding the emergence of the ``ferromagnetic'' phase, in which all velocities are aligned \cite{Toner2005}.
On the other hand, systems that have no net motion can still have a lot of internal motion, but characterizing the dynamics as collective is more difficult.
Based on studies of insect swarms, the new paradigm to define collective motion is through correlations \cite{Ni2015}.
In midge swarms, for example, insects appear to be moving at random, but careful examination of trajectories reveal strong correlations at various scales \cite{Puckett2015,Cavagna2017}.
The dense sheep flock presented in this paper is very different from gas-like insect swarms, and while its center of mass is stationary or slow-moving, it exhibits strong velocity correlations with interesting properties.
These correlations exist at both short and long range, spanning the entire flock, but also avoiding its interior by localizing on the edge (Fig.~\ref{figCorr}, Fig.~\ref{figSheepDog}c,d).
They exist in different behavioral regimes (Phase~1 and Phase~2), persist over time, and synchronize with an external stimulus, the dog (Fig.~\ref{figSheepDog}e,f).
These characteristics, rarely observed, are probably a consequence of the 
%liquid-like state 
strong cohesion of the flock, which propagates changes of velocity quickly.
An interesting extension of these observations would be to determine whether these correlations can be modeled as viscous (or viscoelastic) effects.

% add plosone paper?

%liquid jet
%correlations

\subsection{Edge dynamics}
A second important observation is the propensity of the sheep flock, under stress, to adopt and maintain a circular shape and dense configuration (Fig.~\ref{figDischarge}).
This is visible both in Phase~1 and Phase~2.
When perturbations force departure from the circular shape, the natural tendency is to go back to it as seen in the flock shape and divergence field in Fig.~\ref{fig2}.
This finding is not novel; the behavioral reasons seem straightforward, as sheep at the edge are more at risk of being attacked by a predator, therefore the flock overall minimizes its perimeter \cite{Hamilton1971,Ginelli2015}.
In this paper, we also report new characterizations that, if confirmed at a broader scale, could help refine flocking models in the continuum limit.
First, while the interior of the flock remains stationary, significant motion happens at the edges.
Second, in certain instances, the motion at the edge propagates in a wave-like manner.
We hypothesize that edge motion is present because sheep on the edge are the only ones that: 1) have the ability to move; 2) receive visual clues about their environment and external stimuli, like dog commands. 
These observations reveal that edge dynamics can significantly affect the dynamics of the entire flock.
Indeed, as sheep move along the edge, new sheep become exposed to the edge, which together can create a net motion of the flock (see for example Fig~\ref{figDischarge}).
In addition, interference of edge waves can disrupt and fragment the flock, as seen at the end of Phase~1 (see Fig.~\ref{fig2}).

\subsection{Flock cohesion}
Another striking feature of the sheep flock studied here is its strong cohesiveness, both at small (Phase~1) and large (Phase~2) average velocity.
Together with the edge dynamics described above, these observations show that models for the collective behavior of dense sheep flocks need to incorporate strong cohesive (attractive) terms, that also properly account for edge dynamics.
In the past, different paradigms have been proposed.
In swarms of midges and other disperse insects, where individuals are far apart, the model of a confining potential has often been suggested \cite{Puckett2015}, notably to account for the constant, finite size of the swarm.
In swarms of ants and bees, where individuals grasp on each other, the model of emergent surface tension has been invoked, in analogy with the behavior of liquids \cite{Hu2016,Peleg2018,Smith2019}.
Surface tension can explain, for example, the shape and flow properties of these aggregates.
Both models, emergent central potential and emergent surface tension, are consistent with our main findings, namely minimization of flock perimeter length, and propagation of edge waves.
However, due to the high density and cohesiveness of sheep flocks, the surface tension model might be more adequate, especially in a continuum framework, although, importantly, sheep to not physically pull onto each other.
Further experiments might be able to refine the description of boundary terms, for example by measuring precisely the dispersion relation of edge waves.

%%Our study, however, presents new characterizations, and two important observations.
%%First, the edge of the flock is able to propagate waves (Fig.~\ref{figWave}).
%%Second, the pressure inside the flock is inversely proportional to some (increasing) function of the flock radius $R$ (Fig.~\ref{figDischarge}d).
%%Together, these results are all consistent with a simple property of liquids, namely the surface tension $\sigma$, whereby a small change in surface area $dA$ costs a small change in energy $dE = \sigma dA$.
%%One major implication of surface tension is that a volume of fluid minimizes its overall surface area; in the absence of constraints, the corresponding shape is spherical (or circular in 2D).
%%In addition, surface tension creates an excess pressure $P$ in a drop of radius $R$, $P = \sigma/R$ (Laplace pressure).
%%Finally, surface tension permits the propagation of capillary waves, which in the absence of external forces, have dispersion relation $\omega^2 \sim q^3$.
%%Interestingly, the data in Fig.~\ref{figWave}f also shows an approximately linear relation between $\omega$ and $q_\theta$.
%The data in Fig.\ref{figWave}f is consistent with this relation, but insufficient to conclude definitively.
%By further characterizing the edge dynamics, it might be possible to qualify

% circular shape
% wave propagation
% pressure

\section{Conclusion}

The sheep flock analyzed in this paper switches between two distinct behavioral modes.
In the first phase, the flock has no net motion, and develops fluctuations in shape and velocity distributions. 
However, unlike midge swarms, which are also stationary, the flock remains dense, so that motion is constrained and visual clues are limited to immediate vicinity.
Consequently, sheep do not move at random, and most of the motion happens at the edges.
Edge dynamics therefore appears as a fundamental aspect of flock behavior.
%, which as far as we know is a fairly unexplored concept.
Under excitation from an external, or internal, agent, fluctuations on the edge propagate in a wave-like manner, where individual sheep move very little compared to the excursion of the deformations.
Possibly because these fluctuations add up, the flock exhibits strong (negative) correlations that span its entire length.
These correlations are considered a strong signature of emergent collective behavior.

Velocity correlations across the flock are also seen in the second phase, where the flock is polarized and flows through a gate, although these correlations are then positive.
The dynamics shows less fluctuations.
The fact that the same flock exhibits different collective dynamics at different stages in the movie is further evidence that environmental contextual information conditions collective behavior \cite{Ni2015}.
It will be interesting to investigate further what dynamical patterns are conserved across different behavioral modes.

Many observations described in this paper, such as the propensity of the flock to remain in a circular shape and the propagation of waves along the edge, highlight the importance of models to properly account for cohesion and surface dynamics, through for example a confining potential or a line tension.
%hint at the existence of a line tension, analogous to interfacial tension in fluids.
The hypothesis of a line (or surface) tension in dense flocks is appealing notably because it could be easily incorporated into hydrodynamic models of swarms.
While surface tension is a known effect in compact swarms of ants or bees \cite{Hu2016,Peleg2018}, it is important to note that in these systems, insects are in contact and pulling onto each other, therefore capable to exert direct mechanical forces (traction) between each other, which is not the case in the sheep flock \cite{Smith2019}.
Any line tension in a sheep flock would be purely behavioral.
While our analysis gives some evidence towards it, further experiments are needed to carefully assess the validity and origin of a possible emergent line tension in dense flocks.

%\section{Ethical statement}
%The authors do not condone any sort of abuse or violence against any animal.

%\section{Acknowledgments}
\ack
We are very grateful to David Hu for encouraging this project from the beginning, and to Michele Ferraro for helping us interpret the dynamics in the video and providing general information on sheep and shepherding dogs.
We would like to thank Orit Peleg, Gary Nave, and anonymous reviewers for valuable feedback, and Sophie Backer and \'{E}lie Sarfati for helpful suggestions.
Finally, we thank the unknown author of the sheep video.

% interactions topological rather than metric [Bialek PNAS 2012]
% groups that don't display net motion
% models less successful with aggregations that don't diplay net motion Ouellette pramana
% importance to study fluctuations Ouellette pramana
% continuum model treats all animals as identical, no heterogeneity, no spontaneous "leader"
% attraction to, repulsion from, alignment with neighbours are emergent epiphenomena rather than low-level interactions Ouellette Pramana

\section{Supplemental Material}
In the Supplemental Material, we provide additional details on the processing methods and a short discussion on density fields.
We also supply the original movie (Movie S1), and the same movie with overlaid metrics: the divergence field (Movie S2), the vorticity field (Movie S3), and representative trajectories (Movie S4) with some visual adjustments, to avoid overcrowding.
Some of the MATLAB code used for this paper is available at the link in Ref.~\cite{GitHub}.

\section{References}

\bibliography{iopart-num}

\end{document}